\documentclass[twocolumn,prb]{revtex4}

\usepackage{epsf}

\begin{document}
%\draft
\preprint{}
\title{Hole-doping effects on a frustrated spin ladder}
\author{Akira Kawaguchi, Akihisa Koga,
Kouichi Okunishi$^1$ and Norio Kawakami}
\address{Department of Applied Physics,
Osaka University, Suita, Osaka 565-0871, Japan \\ 
$^1$Department of Physics, Niigata University, 
Igarashi 2, Niigata 950-2181 Japan}
\date{\today}
\begin{abstract}
Hole-doping effects are investigated on the {\it t-J} ladder model
with the linked-tetrahedra structure. 
We discuss how a metal-insulator transition occurs upon 
hole doping with particular emphasis on the effects of 
geometrical frustration. By computing the electron density 
and  the spin correlation function 
by the density matrix renormalization group, 
we show that strong frustration triggers
a first-order transition to a metallic phase,
when holes are doped into the plaquette-singlet phase. 
By examining spin excitations in a metallic case
in detail, we discuss whether the spin-gap phase persists
upon hole doping according to the strength of frustration.
It is further shown that  the lowest excited state in a 
spin-gap metallic phase can be 
described in two independent quasiparticles. 
\end{abstract}


\maketitle

%%%%%%%%%%%%%%%%%%%%%%%%%%%%%%%%%%%%%%%%%%%%%%%%%%%%%%%%%%%%%%%%%%%
%%%%%%%%%%%%%%%%%%%%%%%%%%%%%%%%%%%%%%%%%%%%%%%%%%%%%%%%%%%%%%%%%%%
%%%%%%%%%                   　　　　  %%%%%%%%%%%%%%%%%%%%%%%%%%%%%
%%%%%%%%% 1.  Introduction  　　　　  %%%%%%%%%%%%%%%%%%%%%%%%%%%%%
%%%%%%%%%                   　　　　  %%%%%%%%%%%%%%%%%%%%%%%%%%%%%
%%%%%%%%%%%%%%%%%%%%%%%%%%%%%%%%%%%%%%%%%%%%%%%%%%%%%%%%%%%%%%%%%%%
%%%%%%%%%%%%%%%%%%%%%%%%%%%%%%%%%%%%%%%%%%%%%%%%%%%%%%%%%%%%%%%%%%%

\section{Introduction}

Quantum spin systems with geometrical frustration have attracted 
considerable attention recently. There are a number of typical 
materials, such as  the spin-gap compound 
SrCu$_2$(BO$_3$)$_2$ \cite{Kageyama}
with the orthogonal-dimer structure,
 \cite{Shastry,Miyahara,Koga_let} the pyrochlore compounds  with
the lattice of corner-sharing tetrahedral network of 
magnetic ions,
 \cite{Harris,Canals,Isoda,Koga,Tsunetsugu,Ueda,Fujimoto} etc.

Besides  frustrated spin systems studied intensively so far, 
metallic systems with geometrical frustration have 
attracted much interest. For instance, it has been claimed
 that frustration due to a tetrahedral 
network of V ions may be important to understand heavy-fermion behavior of
the metallic pyrochlore compound  $\rm LiV_2O_4$. \cite{LiV2O4}  
Also, another prototypical pyrochlore
compound  $\rm Y(Sc)Mn_2$ \cite{Y(Sc)Mn2} shows spin-liquid
behavior in a metallic phase.

Stimulated by these experimental findings, we  study here 
the hole-doping effects on fully frustrated spin systems.
To this end, we  consider  a simplified model
in one dimension (1D), which still possesses the 
essence of strong  frustration in the above-mentioned systems.  
To be precise, we employ the {\it t-J} ladder model with 
the orthogonal dimer structure, which also 
has linked-tetrahedra as a key structure. 
The advantage to employ the 1D model is that this simplification 
allow us to accurately calculate physical quantities 
with use of such a reliable numerical method 
as the density matrix renormalization group (DMRG),\cite{White}
%%allows us to precisely calculate physical quantities, 
and thus precisely study 
  the effects of frustration on 
hole-doped systems. 

This paper is organized as follows. 
In Sec. II, we introduce the  frustrated {\it t-J} 
ladder model, and then in Sec. III
study a metal-insulator transition upon  hole doping. 
By computing spin correlation functions, we  
clarify the origin of the first-order phase transition found there. 
In Sec. IV, spin excitations in a metallic phase are discussed in
detail.  It is clarified to what extent a spin-gap metallic 
phase persists around the insulating phase. In particular, 
we show that a spin triplet excitation in the spin-gap phase 
can be described in terms of two independent quasiparticles. 
We obtain the 
phase diagram in Sec. V, and summary and discussions are  given in Sec. VI.

%%%%%%%%%%%%%%%%%%%%%%%%%%%%%%%%%%%%%
\section{Frustrated Ladder Model}
%%%%%%%%%%%%%%%%%%%%%%%%%%%%%%%%%%%%%
%%%%%%%%%%%%%%%%%%%%%%%%%%%%%%%%%%%%%

\begin{figure}[htb]
\begin{center}
\vspace{-0.cm}
\hspace{-3.0cm}
\leavevmode \epsfxsize=150mm
\epsffile{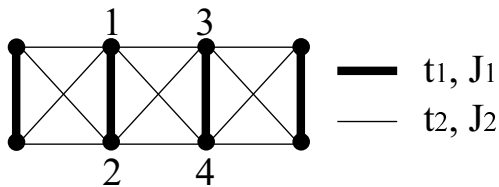}
\vspace{-7.0cm}
\end{center}
\caption{ Frustrated ladder system, which is often referred to as
a linked-tetrahedra chain.
The thick and thin lines
correspond to the parameters  $J_1$ ($t_1$) and  $J_2$ ($t_2$),
respectively.
}
\label{model1}
\end{figure}

The model we treat here is a frustrated electron ladder
which is an extension of the  
spin-1/2 Heisenberg  model with the orthogonal-dimer structure 
introduced by Gelfand. \cite{Gelfand}
The model is illustrated schematically  
 in Fig. \ref{model1}. It has two competing 
antiferromagnetic interactions $J_1$ and $J_2$, giving rise to 
strong  geometrical frustration.
In a hole-doped case, we use the {\it t-J} model where   
electron hopping is assumed to  have two distinct values,
 as shown in Fig.  \ref{model1}.
The Hamiltonian is thus given by
%%%%%%%%%%%%%%%%%%%%%%%%%%%%%%%%%%%%%%%%%%%%%%%%%%%%%%%%%%%%%%%%%%%%%%%%%%
\begin{eqnarray}
%{\cal H}&=&
H&=&
    -\sum_{(i,j),\sigma} t_{\alpha}
              c^{\dag}_{i\sigma} c_{j\sigma}
    +\sum_{(i,j)} J_{\alpha}
          \left(  {\bf S}_i\cdot{\bf S}_{j}-\frac{1}{4} n_i n_j \right),
%
%%\nonumber \\
%%  &&-t_{2}\sum_{<i,j>,\sigma}
%%             c^{\dag}_{i\sigma} c_{j\sigma}
%%    +J_{2}\sum_{<i,j>}
%%          \left(  {\bf S}_i\cdot{\bf S}_{j} -\frac{1}{4} n_i n_j \right)
%
%%\nonumber \\
%%%       &&  - \mu{\sum_{i}}n_{i}
%
\label{TJ_EQ}
\end{eqnarray}
%%%%%%%%%%%%%%%%%%%%%%%%%%%%%%%%%%%%%%%%%%%%%%%%%%%%%%%%%%%%%%%%%%%%%%%%%%
where the summation is taken over nearest 
neighbor bonds ($i$,$j$) indicated 
 by the thick and thin lines in  Fig. \ref{model1}.
Here, $c_{i\sigma}$ is the 
annihilation operator of an electron with 
spin $\sigma$ at the $i$-th site and
$n_{i}$ ($S_{i}$) is the electron number (spin-1/2) operator. 
As mentioned above, hopping $t_i$ (exchange coupling $J_i$) takes 
either $t_1$  or $t_2$ ($J_1$ or $J_2$). 
Note that doubly occupied state at each site are implicitly forbidden in the 
{\it t-J} model.

%%%%%%%%%%%%%%%%%%%%%%%%%%%%%%%%%%%%%%%%%%%%%%%%%%%%%%%%%%%%%%%%%%%%%%%
%%%%%%%%%%%%%%%%%%%%%%%%%%%%%%%%%%%%%%%%%%%%%%%%%%%%%%%%%%%%%%%%%%%

We note here that the above model \cite{Gelfand,Honecker,Brenig,Kim} 
has several remarkable properties in addition to those for ordinary  
1D frustrated spin systems: (i) it is 
a 1D analogue of the 2D orthogonal-dimer
model (Shastry-Sutherland model \cite{Shastry,Miyahara}) 
relevant for SrCu$_2$(BO$_3$)$_2$, for which 
the ground state is exactly given by
a direct product of decoupled dimers; (ii) it is also regarded as
a 1D variant of the pyrochlore system 
(so-called linked-tetrahedra chain), which has frustrated 
 tetrahedra as a key structure. \cite{Gelfand}
A remarkable point related to (ii) 
is that, for noninteracting electrons on this lattice, 
the model has a dispersionless flat-band mode, 
%%%%%%%%%%%%%%%%%%%%%%%%%%%%%%%%%%%%%%%%%%%%%%%%%%%%%%%%%%%%%%%%%%%%%%%%%%
\begin{eqnarray}
\varepsilon (k)=\left\{
 \begin{array}{l}
   t_1 \\
   -t_1-4t_2\cos(k),
 \end{array}
  \right.
\label{band}
\end{eqnarray}
%%%%%%%%%%%%%%%%%%%%%%%%%%%%%%%%%%%%%%%%%%%%%%%%%%%%%%%%%%%%%%%%%%%%%%%%%%
which reflects  special geometry of the system, 
as is the case for the 3D pyrochlore lattice. 
Therefore, using this simplified model
we can also explore  the role of a flat-band mode, 
which is important to understand 
properties of the pyrochlore system
near half filling.

%%For $t_2/t_1 < 0.5$, the two bands are completely separated, 
%%while for $t_2/t_1 > 0.5$,  the upper flat-band  is 
%%in the lower band.  

%%%%%%%%%%%%%%%%%%%%%%%%%%%%%%%%%%%%%%%%%%%%%%%%%%%%%%%%%%%%%%%%%%%
%%%%%%%%%%%%%%%%%%%%%%%%%%%%%%%%%%%%%%%%%%%%%%%%%%%%%%%%%%%%%%%%%%%
\section{Metal-Insulator transition}
%%%%%%%%%%%%%%%%%%%%%%%%%%%%%%%%%%%%%%%%%%%%%%%%%%%%%%%%%%%%

%%%%%%%%%%%%%%%%%%%%%%%%%%%%%%%%%%%%%%%%%%%%%%%%%%%%%%%%%%%%%%%%%%%%%%%%%%%%
\begin{figure}[htb]
\begin{center}
\vspace{-0.cm}
\hspace{0.0cm}
\leavevmode \epsfxsize=90mm 
\epsffile{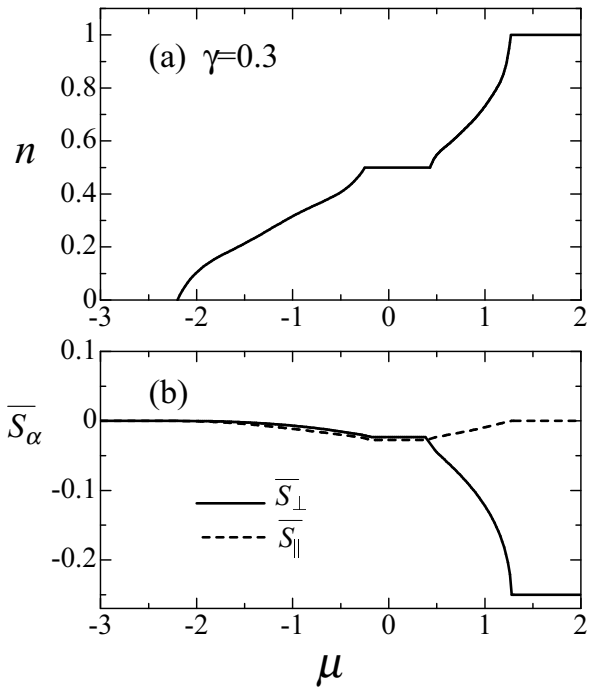}
\vspace{-3.5cm}
\end{center}
\caption{ (a) Electron density $n$ and 
(b) short-range spin correlations $\bar{S}_{\alpha}$ 
as a function of the chemical potential 
$\mu$ for $\gamma=0.3$ at $J_1/t_1=0.25$. 
The system is in a dimer-singlet insulating phase at 
half filling $n=1$.
}
\label{mu_n_03}
\end{figure}
%%%%%%%%%%%%%%%%%%%%%%%%%%%%%%%%%%%%%%%%%%%%%%%%%%%%%%%%%%%%%%%%%%%

%%%%%%%%%%%%%%%%%%%%%%%%%%%%%%%%%%%%%%%%%%%%%%%%%%%%%%%%%%%%%%%%%%%%%%%%%%%%
\begin{figure}[htb]
\begin{center}
\vspace{-0.cm}
\hspace{0.0cm}
\leavevmode \epsfxsize=90mm 
\epsffile{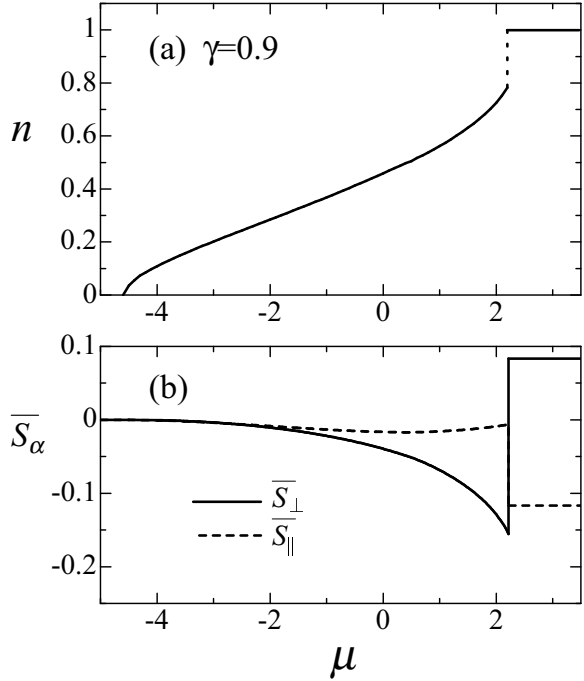}
\vspace{-3.5cm}
\end{center}
\caption{(a) Electron density $n$ and 
(b) short-range spin correlations $\bar{S}_{\alpha}$ 
as a function of the chemical potential 
 $\mu$ for $\gamma=0.9$ and $J_1/t_1=0.25$.
This corresponds to hole doping in the plaquette phase.
The dotted line in (a) indicates a jump 
associated with a first-order transition. 
}
\label{mu_n_09}
\end{figure}
%%%%%%%%%%%%%%%%%%%%%%%%%%%%%%%%%%%%%%%%%%%%%%%%%%%%%%%%%%%%%%%%%%%

Let us start our  discussions with the ground-state properties of 
the {\it t-J} model in a metallic phase. We particularly
focus on the nature of a metal-insulator transition induced by hole doping.
Here, we use a variant of the infinite DMRG method, so-called
 product-wave-function renormalization group 
method \cite{Nishino}, to compute several quantities in a bulk limit. 

In the undoped spin model, the system  
is either in the exact dimer-singlet phase ($J_2/J_1<0.71$)
or the plaquette-singlet phase ($J_2/J_1>0.71$).\cite{Gelfand}
We thus discuss the hole-doping effects on
these phases separately.  In the following analyses,
we set $t_2/t_1 = J_2/J_1 \equiv \gamma$,
and mainly show the results of $J_1/t_1=0.25$ ($t_1=1.0$) 
for simplicity. 

%%%%%%%%%%%%%%%%%%%%%%%%%%%%%%%%%
\subsection{dimer phase}
%%%%%%%%%%%%%%%%%%%%%%%%%%%%%%%%

We  begin with the 
dimer phase, which is stabilized at half filling for the ratio of the 
exchange couplings,  $\gamma < 0.71 $. 
The calculated  electron density $n$ is shown in Fig. \ref{mu_n_03}(a)
as a function of the chemical potential $\mu$.

It is seen that hole-doping smoothly 
drives the system to a metallic phase. 
At quarter filling ($n=1/2$), there appears
a plateau in $n$, which is due to 
formation of a CDW insulating state.
It is instructive to compare this smooth transition  
with first-order transitions found  in the  magnetization 
curve at half filling
in the dimer phase: \cite{Honecker}
the magnetization exhibits jumps  among
the plateaus at zero, half, and full magnetization.
The discontinuity in the 
magnetization reflect the fact that a triplet excited state
 in the  dimer phase is
completely localized due to the orthogonal-dimer structure. 
This is not  the case for hole doping: 
doped holes are indeed heavy but still mobile, allowing
a smooth transition to the metallic phase.

We next compute spin correlation functions, which are defined 
in the rung ($\perp$) and chain ($\parallel$) directions as,
%%%%%%%%%%%%%%%%%%%%%%%%%%%%%%%%%%%%%%%%%%%%%%%%%%%%%%%%%%%%%%%%%%%%%%%%%%
\begin{eqnarray}
\bar{S}_{\perp}&=& \langle S^z_1 S^z_2 \rangle, 
\nonumber \\
\bar{S}_{\parallel} &=& \langle S^z_1 S^z_3 \rangle 
                    (=\langle S^z_1 S^z_4 \rangle ). 
%     \frac{1}{4}
%     ( <S^z_1 S^z_3>+<S^z_1 S^z_4> 
%\nonumber \\
%  && 
%     +<S^z_2 S^z_3>+<S^z_2 S^z_4>   ). 
\nonumber
\label{TJ_EQ}
\end{eqnarray}
%%%%%%%%%%%%%%%%%%%%%%%%%%%%%%%%%%%%%%%%%%%%%%%%%%%%%%%%%%%%%%%%%%%%%%%%%%
%We make use of the infinite DMRG method
%to compute these quantities in a bulk limit.  
Shown  in Fig. \ref{mu_n_03}(b) are the computed spin correlation functions.
As the system approaches half filling
from a metallic side,  $\bar{S}_{\perp}$ along the 
rung  rapidly decreases down to $-1/4$
whereas  $\bar{S}_{\parallel}$  along the chain 
becomes almost zero.  Recall  that the values of
$\bar{S}_{\perp}=-1/4$ and $\bar{S}_{\parallel}=0$ 
are expected  for a system with isolated dimers. 
Therefore, the above results of correlation functions imply that 
the system near half filling  is  regarded  
as a resonating metallic state of decoupled dimers.

%%%%%%%%%%%%%%%%%%%%%%%%%%%%%%%
\subsection{plaquette phase}
%%%%%%%%%%%%%%%%%%%%%%%%%%%%%%%

When holes are doped into the plaquette-singlet phase,
the nature of a metal-insulator transition is completely changed. 
Shown in Fig. \ref{mu_n_09}(a) is the electron density 
as a function of the chemical potential. 
%when holes are doped  in the  plaquette-singlet phase.  
It is remarkable that 
the system is driven to a metallic phase via 
a {\it first-order} phase transition, 
which is characterized by a jump in the 
electron  density.
%%%By carefully treating the system  by the DMRG method, 
%%%we find the hysteresis characteristic of the first-order 
%%%phase transition.  Namely, the broken lines in the figure
%%%represent the results calculated  
%%%for  increasing $\mu$  (right arrow) 
%%%and decreasing $\mu$ (left arrow). 

We can see the origin of the first-order metal-insulator transition 
by checking short-range spin correlation functions
in the rung as well as leg directions,
shown in Fig. \ref{mu_n_09}(b). At half filling, 
$\bar{S}_{\perp}>0$ and $\bar{S}_{\parallel}<0$, 
being consistent with the plaquette-singlet ground state.
However, once the system is driven to a metallic phase 
via the first-order transition, both of 
$\bar{S}_{\perp}$ and $\bar{S}_{\parallel}$ become
negative. Furthermore, when the system approaches half-filling 
from a metallic side,
 $\bar{S}_{\perp}$ decreases with negative values 
while $\bar{S}_{\parallel}$ gets very small.
Namely, in a metallic phase close to half filling 
 dimer correlations along the rung are largely enhanced. 
Hence, in the vicinity of half filling, 
this dimer-dominant state strongly competes
with the plaquette-singlet state energetically, 
giving rise to  the first-order phase transition
observed above.

We note here that such a first-order metal-insulator transition
only occurs when holes are doped in 
the  plaquette phase, but not in the
dimer phase. It is also instructive to compare this with 
the results known for the 
magnetization curve at half filling, where 
first-order phase transitions  occur only in the dimer phase
($\gamma <0.71$). \cite{Honecker}
These apparently different aspects
in phase transitions come from  the fact that 
hole doping induces a competition between the dimer- and 
plaquette-singlet states, whereas a magnetic field induces 
a competition between the dimer- (or plaquette-)
singlet state and the triplet state.

%%%%%%%%%%%%%%%%%%%%%%%%%%%%%%%%%%%%%%%%%%%%%%%%%%%%%%%%%%%%%%%%%%%
%%%      SS. 3 
%%%%%%%%%%%%%%%%%%%%%%%%%%%%%%%%%%%%%%%%%%%%%%%%%%%%%%%%%%%%%%%%%%%

%%%%%%%%%%%%%%%%%%%%%%%%%%%%%%%%%%%%%%%%%%%%%%%%%%%%%%%%%%%%%%%%%%%%%%%
\begin{figure}[bth]
\begin{center}
\vspace{-0.cm}
\hspace{0.0cm}
\leavevmode \epsfxsize=90mm 
\epsffile{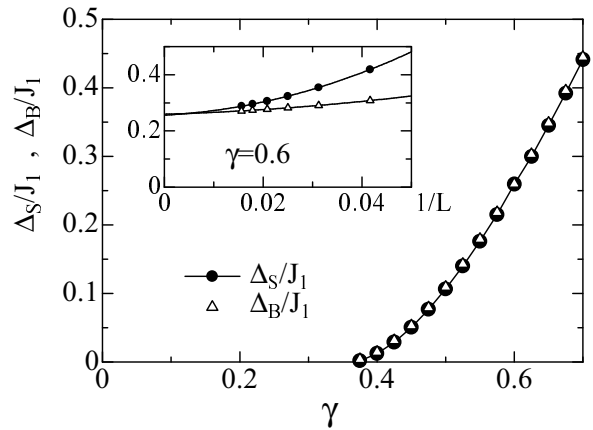}
\vspace{-7.0cm}
\end{center}
\caption{ Spin gap $\Delta_S$ and binding 
energy $\Delta_B$ of two holes ($N_h=2$) for $J_1/t_1=0.25$. 
Inset shows a finite-size scaling analysis of 
$2 \times L$ systems for $\gamma=0.6$.}
\label{D_Gap}
\end{figure}
%%%%%%%%%%%%%%%%%%%%%%%%%%%%%%%%%%%%%%%%%%%%%%%%%%%%%%%%%%%%%%%%%%%

%%%%%%%%%%%%%%%%%%%%%%%%%%%%%%%%%%%%%%%%%%%%%%%%%%%%%%%%%%%%%%%%%%%%%%%
\begin{figure}[htb]
\begin{center}
\vspace{-0.cm}
\hspace{-0.5cm}
\leavevmode \epsfxsize=90mm 
\epsffile{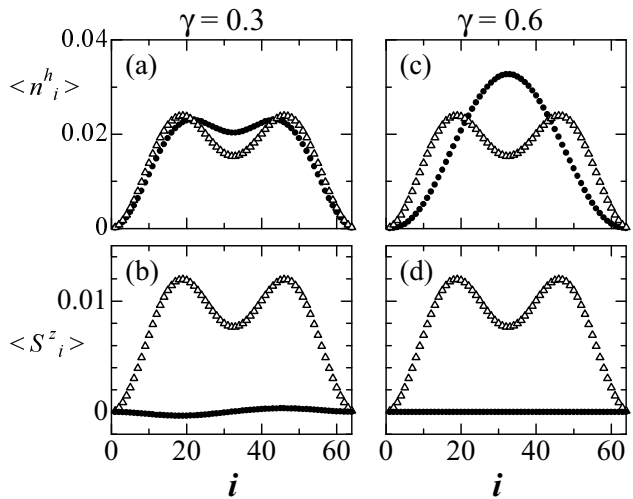}
\vspace{-6.0cm}
\end{center}
\caption{ Hole density $\langle n^h_i \rangle$ and spin density 
$\langle S^z_i \rangle $ 
for two holes on the $2\times 64$ system:
 $\gamma=0.3$ for (a) and (b) while $\gamma=0.6$
for (c) and (d). 
Filled circles and open triangles correspond to 
the ground state and excited states, respectively.
}
\label{Ni_Sz}
\end{figure}
%%%%%%%%%%%%%%%%%%%%%%%%%%%%%%%%%%%%%%%%%%%%%%%%%%%%%%%%%%%%%%%%%%%

%%%%%%%%%%%%%%%%%%%%%%%%%%%%%%%%%%%%%%%%%%%%%%%%%%%%%%%%%%%%%%%%%%%%%%%
\begin{figure}[htb]
\begin{center}
\vspace{-0.cm}
\hspace{-0.0cm}
\leavevmode \epsfxsize=90mm 
\epsffile{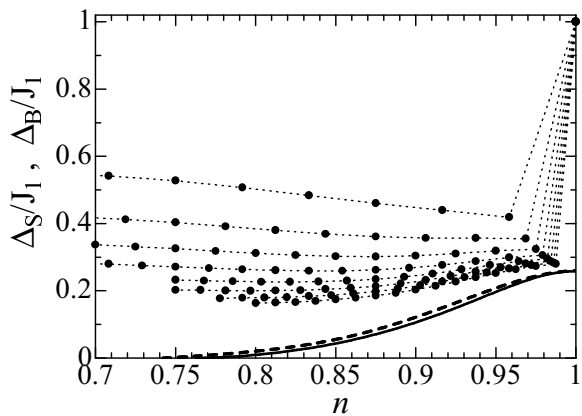}
\vspace{-7.0cm}
\end{center}
\caption{ Spin gap (solid line) and binding energy (broken line) as a 
function of $n$ ($=1-N_h/L$):
$\gamma=0.6$ and  $J_1/t_1=0.25$. 
The dotted lines with  filled circles are the results of $\Delta_S $ for
$L=24,32,40,48,56,64,72,80$ from top to bottom. 
}
\label{Sgapf}
\end{figure}
%%%%%%%%%%%%%%%%%%%%%%%%%%%%%%%%%%%%%%%%%%%%%%%%%%%%%%%%%%%%%%%%%%%

%%%%%%%%%%%%%%%%%%%%%%%%%%%%%%%%%%%%%%%%%%%%%%%%%%
\section{Spin excitations in a metallic phase}
%%%%%%%%%%%%%%%%%%%%%%%%%%%%%%%%%%%%%%%%%%%%%%%%%%

We now discuss spin excitations in a hole-doped system near half filling. 
We start with low-energy excitations in a system with two holes. 
As mentioned above, the system is driven to a metallic phase immediately
upon hole doping  for the dimer phase ($0< \gamma <0.71$).
On the other hand, for the plaquette phase ($\gamma >0.71$), 
 doped holes may be in a localized state, as a result of  
 a first-order transition.  Namely,  holes are accommodated in a degenerate 
dispersionless level up to a critical hole density.
In the following, we thus pay our attention to the former region, 
$0< \gamma <0.7$, which is more interesting as far as the system with
a small number of  holes is concerned.

In order to discuss excitations in  a doped case, we consider
two quantities, i.e. 
the spin gap $ \Delta_S (N_h)$ and the binding energy 
 $ \Delta_B (N_h)$ respectively defined for $N_h$ (even number) holes 
on $2\times L$ lattice sites as,  
%%%%%%%%%%%%%%%%%%%%%%%%%%%%%%%%%%%%%%%%%%%%%%%%%%%%%%%%%%%%%%%%%%%%%%%%%%
\begin{eqnarray}
\Delta_S (N_h)&=& E_0 (N_h ,S^z=1)-E_0 (N_h ,S^z=0) , 
 \\
\Delta_B (N_h)&=& 2 E_0 (N_h-1,S^z=1/2)-E_0 (N_h ,S^z=0) 
\nonumber\\
              &&-E_0 (N_h-2 ,S^z=0). 
\label{Eq_D_s}
\end{eqnarray}
%%%%%%%%%%%%%%%%%%%%%%%%%%%%%%%%%%%%%%%%%%%%%%%%%%%%%%%%%%%%%%%%%%%%%%%%%%
In Fig. \ref{D_Gap}, we show the spin gap $\Delta_S$ 
and the binding energy $\Delta_B$ computed for
 two doped holes ($N_h=2$). 
The inset shows a finite-size scaling of $\Delta_S$ 
and $\Delta_B$ for $\gamma=0.6$. 
It should be noticed that 
the spin gap and the binding energy monotonically increase  
as $\gamma$ increases from $\gamma\simeq 0.35$. 
It is seen that the spin gap $\Delta_S $ just 
coincides with the binding energy $\Delta_B $, implying 
 that a pair-breaking excitation in a metallic phase 
 may be described by two independent quasiparticles. 
It is to be noted that the above spin gap in a doped system
 is totally different from the spin gap defined
at half filling, $\Delta_S = J_1$, as we will see
below.

Shown  in Fig. \ref{Ni_Sz} is a spatial distribution of the hole 
density $\langle n^h_i \rangle$ ($n^{h}_i \equiv 1-n_i$)
and the spin density $\langle S^z_i \rangle$ 
for two holes on a $L=64$ system.
Here filled circles and  open triangles show the results for 
the singlet ground state
and  for a triplet excited state, respectively. 
For $\gamma=0.6$, as shown by  filled circles in Fig. \ref{Ni_Sz}(c), 
a hole-pair exists in the ground state, implying the 
existence of an attractive interaction between two holes in the ground state. 
For a triplet  excitation, the bound state is 
broken and two quasiparticles appear, 
as seen from open triangles in Fig. \ref{Ni_Sz}(c) and (d). 
On the other hand, the situation is somewhat different for $\gamma=0.3$, where 
 the repulsive interaction seems to exist even in the ground state,
 as shown by  filled circles in Fig. \ref{Ni_Sz}(a) and (b). 
We have checked that a repulsive interaction in the latter case is 
mainly caused by finite-size effects. Therefore,
it is expected that such a repulsive interaction 
 becomes small for large $L$, and in the thermodynamic limit, 
 doped holes may behave as two independent quasiparticles 
in the case of $\gamma=0.3$. 

We now turn to the case with a finite density of holes. 
In Fig. \ref{Sgapf}, we show the spin gap as a function of  
the electron density $n$ for $\gamma=0.6$. 
The dotted lines with  filled circles are the results of $\Delta_S $  
for finite-size systems. It is seen that the value of the spin gap   
dramatically changes between half-filling and two-hole case,  
and then smoothly changes in the region $N_h \geq 2$. 
We have performed a finite-size scaling analysis of the spin gap 
from the data shown by dotted lines, obtaining 
the curve shown by the  solid line.  Note that the spin gap 
has a discontinuity at half filling, implying that the 
spin gap in a metallic phase has a different origin from 
the half-filled case. Also shown by the broken line is
 the binding energy $\Delta_B$ 
computed  by a similar finite-size scaling analysis. 
Since $\Delta_S$ is nearly equal to $\Delta_B$, 
we can say that  excited states may be described by 
two independent quasiparticles, as we have 
already seen in a system with  two holes. 

\begin{figure}[htb]
\begin{center}
\vspace{-0.cm}
\hspace{0.0cm}
\leavevmode \epsfxsize=90mm 
\epsffile{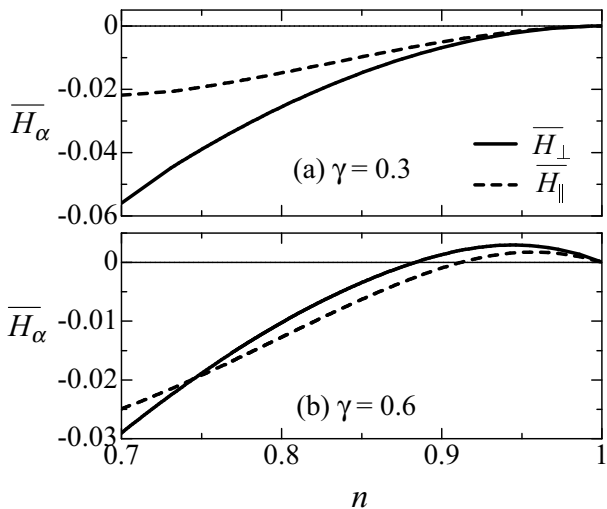}
\vspace{-6.0cm}
\end{center}
\caption{Short-range hole-hole correlations $\bar{H}_{\alpha}$ 
for $\gamma=0.3, 0.6$ and $J_1/t_1=0.25$.
}
\label{Hole_corr}
\end{figure}

In order to see the nature of the
 attractive interaction between holes more clearly, we
compute  short-range hole-hole correlation functions 
$\bar{H}_{\alpha}$  in a bulk limit, which are defined in the rung ($\perp$) and chain 
($\parallel$) directions, 
%%%%%%%%%%%%%%%%%%%%%%%%%%%%%%%%%%%%%%%%%%%%%%%%%%%%%%%%%%%%%%%%%%%%%%
\begin{eqnarray}
\bar{H}_{\perp}&=& \langle n^{h}_1 n^{h}_2 \rangle 
- \langle n^{h}\rangle ^2 , 
\nonumber \\
\bar{H}_{\parallel}  &=& \langle n^{h}_1 n^{h}_3\rangle
-\langle n^{h} \rangle^2 . 
\nonumber
\label{Eq_HH}
\end{eqnarray}
%%%%%%%%%%%%%%%%%%%%%%%%%%%%%%%%%%%%%%%%%%%%%%%%%%%%%%%%%%%%%%%%%%%%%%
In Fig. \ref{Hole_corr}, we show $\bar{H}_{\alpha}$ as
a function of $n$ for $\gamma=0.3$ and $0.6$. For $\gamma=0.6$, 
there is the region with both of $\bar{H}_{\perp}$ and
 $\bar{H}_{\parallel}$ being positive, where 
 an effective interaction between holes is attractive in both
directions.  On the other hand,  
for $\gamma=0.3$,  both of $\bar{H}_{\perp}$ and 
 $\bar{H}_{\parallel}$ are negative
 (i.e. repulsive interaction) in the whole range shown in the figure. 
By comparing these results with Fig. \ref{D_Gap}, 
we see that the region, where holes attract each other, 
 roughly corresponds to the spin-gap metallic region. 
This means that the mechanism of spin-gap formation 
in a metallic phase, which is different from that 
in the insulating phase, is closely related to  the  attractive force  
between holes.

%%%%%%%%%%%%%%%%%%%%%%%%%%%%%%%%%%%%%%%%%%%%%%%%%%%%%%%%%%%%%%%%%%%
%%%%%%%%%%%%%%%%%%%%%%%%%%%%%%%%%%%%%%%%%%%%%%%%%%%%%%%%%%%%%%%%%%%

%%%%%%%%%%%%%%%%%%%%%%%%%%%%%%%%%%%%%%%%%%%%%%%%%%%%%%%%%%%%%%%%%%%
%%%      SS. 3 
%%%%%%%%%%%%%%%%%%%%%%%%%%%%%%%%%%%%%%%%%%%%%%%%%%%%%%%%%%%%%%%%%%%

%%%%%%%%%%%%%%%%%%%%%%%%%%%%%%%%
\section{Phase diagram }
%%%%%%%%%%%%%%%%%%%%%%%%%%%%%%%%

\begin{figure}[htb]
\begin{center}
\vspace{-0.cm}
\hspace{0.0cm}
\leavevmode \epsfxsize=90mm 
\epsffile{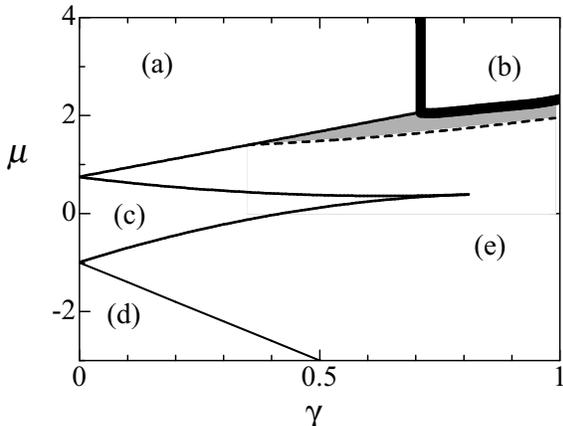}
\vspace{-7.5cm}
\end{center}
\caption{Ground-state phase diagram
for $J_1/t_1=0.25$:
(a) dimer-singlet phase ($n=1$), (b) plaquette-singlet phase ($n=1$), 
(c) CDW phase ($n=1/2$), (d) empty phase ($n=0$), (e) metallic phase, 
and shaded area indicates a spin-gap metallic phase. 
}
\label{PD_1}
\end{figure}

Based on the above results, 
we now obtain the ground-state phase diagram, 
which is  shown in Fig. \ref{PD_1}.
As found  by Gelfand,\cite{Gelfand}  
the ground state at half filling 
 is either in the dimer-singlet (or plaquette-singlet) phase  
with spin gap at half filling for $\gamma < 0.71$ ($\gamma > 0.71$).
These two phases are separated via 
a first-order phase transition when  $\gamma$ is changed.

We have found that the transition from the insulating plaquette phase 
to a metallic phase is of first-order, being contrasted to 
the dimer phase showing a second-order transition.
Furthermore, a spin-gap metallic phase, which is indicated by the 
shaded region in Fig. \ref{PD_1}, 
persists in the vicinity of half-filling, which is indeed caused 
 by  strong frustration.
 For the value $J_1/t_1=0.25$ shown in
 Fig. \ref{PD_1}, we have found that a spin-gap metallic phase appears
in the region of $\gamma > 0.35(\equiv \gamma_S)$. 

When the electron density is further decreased, we encounter
a CDW insulating phase at quarter filling, characterized by
the plateau in $n$-$\mu$ curve. The region of a CDW phase
becomes narrower as $\gamma$ increases, and finally vanishes 
beyond a certain critical value.
As is seen from Fig. \ref{mu_n_03}(b) and Fig. \ref{mu_n_09}(b),
 a metallic phase continuously spreads in 
 the whole region of $\gamma$,
where short-range spin correlations along the rung,
 $\bar{S}_{\perp}$, always take negative values.
This implies that dimer correlations are dominant for
the ground state in a metallic phase, as is usually
the case for a doped antiferromagnetic ladder. 
However, it should be noticed that 
in a metallic phase close to the plaquette-singlet insulator, a
 competition  between the dimer- and plaquette- singlet states 
 may give rise to dual properties in 
physical quantities.

\begin{figure}[htb]
\begin{center}
\vspace{-0.cm}
\hspace{-0.0cm}
\leavevmode \epsfxsize=90mm 
\epsffile{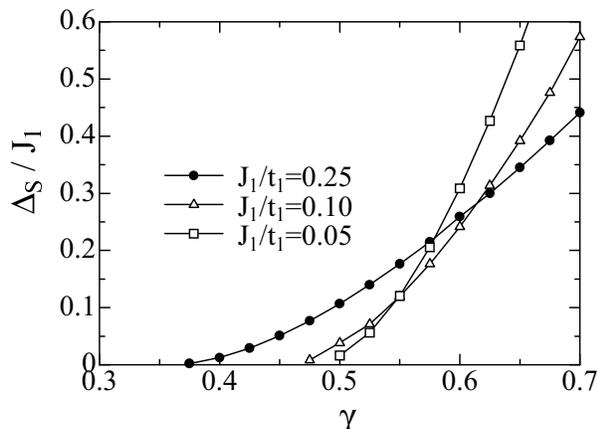}
\vspace{-7.5cm}
\end{center}
\caption{ Spin gap with two holes as a function of $\gamma$ 
for $J_1/t_1=0.25, 0.10, 0.05$. Other parameters are chosen 
as in Fig. \ref{D_Gap}.
}
\label{Sgap}
\end{figure}

We have thus far shown the calculated results by choosing a
specific ratio of hopping and exchange.  It is necessary
to check what is changed when we take other choices of the parameters.
For instance, we show the spin gap of a system with two holes 
in Fig. \ref{Sgap} when the ratio of 
 $J_1/t_1$ is changed.
As $J_1/t_1$ decreases,  the critical point, $\gamma_S$, where the
spin gap starts to develop, is shifted to a 
larger value. Note, however, that
$\gamma_S\rightarrow 0.5$ in 
the limit of $J_1 \ll t_1$. 
Namely, even if $J_1/t_1$ is small, 
a spin-gap metallic phase appears at least in the 
region of $0.5 <\gamma < 0.71$. 
 This feature is different from  the case of the  {\it t-J}
 ladder without frustration,
 \cite{Tune_lad,Dagotto,WhitScala,Siller,Poilblanc,Hayward} 
which has a spin gap only for relatively large $J_1/t_1$.  

%%From the above discussions, we can say that
%%a first-order metal-insulator transition is caused by
%% strong magnetic frustration.
%%Also, the formation of a spin-gap 
%%metallic phase is closely related to frustration.

%We think one of the reasons is the influence of flat-band, Eq. (\ref{band}). 
%A flat-band has a role like phase-separation which indicates 
%the strong attraction,  
%so we think this attraction remains even for $U=\inf$ Hubbard model, 
%$J \ll t$ $t-J$ model. 
%Also $\gamma_S\rightarrow 0.5$ for the limit of $J_1 \ll t_1$  
%associates with 
%a connection of bands structure which a upper flat-band  
%crosses  a lower band for $\gamma=t_2/t_1 > 0.5$. 

%%%%%%%%%%%%%%%%%%%%%%%%  FIGURES  %%%%%%%%%%%%%%%%%%%%%%%%
%%%%%%%%%%%%%%%%%%%%%%%%%%%%%%%%%%%%%%%%%%%%%%%%%%%%%%%%%%%%%%%%%%%

%%%%%%%%%%%%%%%%%%%%%%%%%%%%%%%%%%%%%%%%%%%%%%%%%%%%%%%%%%%%%%%%%%%%%%%
%%%%%%%%%%%%%%%%%%%%%%%%%%%%%%%%%%%%%%%%%%%%%%%%%%%%%%%%%%%%%%%%%%%

%%%%%%%%%%%%%%%%%%%%%%%%%%%%%%%%%%%%%%%%%%%%%%%%%%%%%%%%%%%%%%%%%%%
%%%%%%%%% 4. 　 SUMMARY 　　　　　　  %%%%%%%%%%%%%%%%%%%%%%%%%%%%%
%%%%%%%%%　　　　　　　　　　　　　　 %%%%%%%%%%%%%%%%%%%%%%%%%%%%%
%%%%%%%%%%%%%%%%%%%%%%%%%%%%%%%%%%%%%%%%%%%%%%%%%%%%%%%%%%%%%%%%%%%
\section{Summary and Discussions}

We have studied the hole-doping effects on a 
 spin ladder system with the orthogonal-dimer structure,
 which also possesses linked-tetrahedra as a key structure.
We have investigated  how a metal-insulator
transition is affected by geometrical frustration,
by computing  the electron density and several correlation 
functions by means of the DMRG method.

It has been found that when the system is in the dimer phase, 
hole-doping smoothly drives the system to a metallic phase. 
However, when the system is in the plaquette phase, 
a first-order transition occurs upon hole doping.
The latter first-order transition occurs as a consequence of
a strong competition between the dimer- and  
plaquette-singlet states.  This is different from 
ordinary metal-insulator transitions 
for a frustrated {\it t-J} chain  with next-nearest neighbor
exchange couplings studied so far.\cite{Ogata}

The first-order phase transition found here 
is closely related to unique geometry of the 
linked-tetrahedra lattice,
which causes a dispersionless (spatially isolated)
 mode, as mentioned in the beginning of the paper.
It is to be noted that such a dispersionless mode
caused by special geometry exists for the 3D pyrochlore
lattice  and also for the 2D orthogonal-dimer 
lattice. In this sense,  the present results may 
capture some of characteristic 
properties common in this class of fully  frustrated 
systems with doped holes.  For instance,
a pyrochlore antiferromagnetic system,
indeed  shows a competition between the dimer and plaquette
phases, being similar to the present results. \cite{Koga} 
Therefore, we think that  unusual properties may 
appear in a metallic phase of the doped 
pyrochlore system around the insulating phase.
This is  consistent with the results deduced by
preservative analyses. \cite{Isoda,Fujimoto} 
We also expect that such peculiar properties may emerge even in
2D orthogonal-dimer systems when holes are doped into
an insulating phase around the boundary between the
dimer and plaquette phases. \cite{Koga_let}  Detailed 
theoretical studies on hole-doped 
pyrochlore systems as well as 2D orthogonal-dimer systems are
left as an interesting open problem.

%%%%%%%%%%%%%%%%%%%%%%%%%%%%%%%%%%
\section*{Acknowledgements}
%%%%%%%%%%%%%%%%%%%%%%%%%%%%%%%%%%%%%
This work was partly supported by a Grant-in-Aid from the Ministry 
of Education, Science, Sports and Culture of Japan. 
A part of computations was done at the Supercomputer Center at the 
Institute for Solid State Physics, University of Tokyo
and Yukawa Institute Computer Facility. 
A. Kawaguchi  was supported by 
Japan Society for the Promotion of Science. 
%%%%%%%%%%%%%%%%%%%%%%%%%%%%%%%%%%%%%%%%%%%%%%%

%\clearpage

\end{document}